\documentclass[12pt]{article}\usepackage[]{graphicx}\usepackage[]{color}
\makeatletter
\def\maxwidth{ %
  \ifdim\Gin@nat@width>\linewidth
    \linewidth
  \else
    \Gin@nat@width
  \fi
}
\makeatother

\definecolor{fgcolor}{rgb}{0.345, 0.345, 0.345}

\usepackage{framed}
\makeatletter
\newenvironment{kframe}{%
 \def\at@end@of@kframe{}%
 \ifinner\ifhmode%
  \def\at@end@of@kframe{\end{minipage}}%
  \begin{minipage}{\columnwidth}%
 \fi\fi%
 \def\FrameCommand##1{\hskip\@totalleftmargin \hskip-\fboxsep
 \colorbox{shadecolor}{##1}\hskip-\fboxsep
     \hskip-\linewidth \hskip-\@totalleftmargin \hskip\columnwidth}%
 \MakeFramed {\advance\hsize-\width
   \@totalleftmargin\z@ \linewidth\hsize
   \@setminipage}}%
 {\par\unskip\endMakeFramed%
 \at@end@of@kframe}
\makeatother

\definecolor{shadecolor}{rgb}{.97, .97, .97}
\definecolor{messagecolor}{rgb}{0, 0, 0}
\definecolor{warningcolor}{rgb}{1, 0, 1}
\definecolor{errorcolor}{rgb}{1, 0, 0}
\newenvironment{knitrout}{}{} % an empty environment to be redefined in TeX

\usepackage{alltt}
\renewcommand{\baselinestretch}{1.5}
\usepackage{epsfig}
\usepackage{pstricks}
\usepackage{amsmath}
\usepackage{latexsym}
\usepackage{amssymb}
\usepackage{natbib}
\usepackage{amsthm}
\usepackage{bm}
\usepackage{graphicx}

\theoremstyle{plain}

\theoremstyle{definition}

\IfFileExists{upquote.sty}{\usepackage{upquote}}{}
\begin{document}

\renewcommand{\baselinestretch}{1.5}

\begin{center}
{\large \textbf{The Problem with Assessing Statistical Methods}}
\end{center}
%{\white
%}
\renewcommand{\baselinestretch}{1.0}
\begin{center}
\begin{tabular}{cc}
Abigail A. \ Arnold &Jason L.\ Loeppky  \\[-7pt]
Department of Statistics & Statistics  \\[-7pt]
Western University & University of British Columbia  \\[-7pt]
London, ON N6A 3K7, CANADA & Kelowna, BC V1V 1V7, CANADA  \\[-7pt]
& (jason.loeppky@ubc.ca)
\end{tabular}
\end{center}
%\noindent {\bf \today}

\begin{abstract}
\noindent  In this paper, we investigate the problem of assessing statistical methods and effectively summarizing results from simulations. Specifically, we consider problems of the type where multiple methods are compared on a reasonably large test set of problems. These simulation studies are typically used to provide advice on an effective method for analyzing future untested problems. Most of these simulation studies never apply statistical methods to find which method(s) are expected to perform best. Instead, conclusions are based on a qualitative assessment of poorly chosen graphical and numerical summaries of the results. We illustrate that the Empirical Cumulative Distribution Function when used appropriately is an extremely effective tool for assessing what matters in large scale statistical simulations. 

\medskip

 \end{abstract}
\renewcommand{\baselinestretch}{1.6}

\section{Introduction \label{sect:intro}}

Statistical models are used throughout the sciences and engineering to model complex relationships. Typically, models are chosen on a problem-to-problem basis. While for most examples this is completely justified, there are many cases, especially during model development, where models are being evaluated to provide advice to the practitioner as to what model should be fit.  Obviously there is no one model that will universally outperform the rest.  Recognizing the ``No Free Lunch'' theorem, the logical question to ask is whether one model will perform best over a given class of problems.  Again, we feel that the answer to this question is of course no.  But we do feel that there are certain methods that will have a better chance than other methods.   The current statistics literature is rife with examples of simulations comparing various methods on problems or data.  The presentations of these results often range from tables of results, summarizing means and variances of simulations, to a large number of boxplots or dotplots for various subsets of the data.  

A brief look into recent papers in the Journal of Statistical Computation and Simulation, Journal of Computational and Graphical Statistics, and Canadian Journal of Statistics shows a handful of papers that provide large tables of results or results averaged over many trials that prove it difficult to quickly determine the best method \citep{SheYan2014, YanLvGuo2014,GraApl2015,SchZhaKal2014}. \cite{GraApl2015} apply a Monte Carlo experiment to borehole data and averages the RMSE values for thirty trials for each of the twelve methods being compared.  They find that the  best method is not statistically better than the second best.  The reader of the paper would benefit from seeing a graphical representation of all of the RMSE values to better compare the performance of the two best methods. \cite{MbaPao2014} show improvement by graphically representing their data reasonably well. By averaging values over 50,000 trials, they do not account for variability in their plots. We find that examples like these persist throughout the literature and feel this is problematic.  Motivated by the Benchmarking literature in Optimization this paper argues for the use of the Empirical Cumulative Distribution function (ECDF) to display experimental results which allows the practitioner to quickly assess the appropriateness of a given model.  

This paper is outlined as follows.  In Section \ref{sect:we},  we present a small simulation study related to the choice of experimental design for Computer Experiments and show some summaries of the results.  In Section \ref{sect:sm}, we review standard techniques of comparing optimization algorithms and argue that the same approach should be used to compare statistical methods. In Section \ref{sect:asm} we show various uses of plots adopted from benchmarking to assess the quality of a given statistical method. Finally, we make some concluding remarks in Section \ref{sect:cr}.

\section{Working Example \label{sect:we}} 

As a motivating example we consider the problem of choosing a design for a computer experiment which was disscused in \cite{Gus2012}. Computer models are extensively used to simulate complex physical processes across all areas of science and engineering. In many cases the computer code is slow to run and a statistical model, such as a Gaussian Process (GP), is used to approximate the code \citep{SacWelMit1989,CurMitMor1991,SanWilNot2003}.  Building such a numerical approximation requires the user to specify an experiment design where the code is run and then to build an appropriate GP emulator.

For our working example, we consider some of the tests performed in \cite{Gus2012} to choose an appropriate design for a computer experiment.  The simulation environment selected 2 different versions of 4 generic problems that are evaluated for three different input dimensions.  This is equivalent to having 24 different test problems and each test problem is evaluated for 3 different choices of sample size ($5d,10d,15d$) where $d$ is the number of input dimensions. For a given input design, 50 equivalent designs are constructed by permuting the columns of the original design which is roughly equivalent to 50 replicates of each of the 72 experimental settings.  Each experiment evaluated 7 different experiment designs, which we will label as Methods 1 through 7, where $M1$ is a random latin hypercube (LHS) \citep{McKBecCon1979}; $M2$ is a Maxi-min distance based LHS \citep{MorMit1995}; $M3$ is a zero correlation based LHS \citep{ImaCon1982}; $M4$ is a cosine transformed Maxmin LHS \citep{DetPep2010}; $M5$ is the cosine transformed zero correlation LHS; $M6$ is a space filling Sobol sequence \citep{Sob1967} and finally $M7$ is a simple random sample.
%%Following \cite{LoeSacWel2009} one would expect reasonable performance for problems 
Since the test functions at hand are relatively quick to run, an additional set of test data is used to assess the overall performance of each method. Specifically. model performance is judged using the mean squared error and absolute max error specified as
$$MSE=\sum_{i=1}^m (y_i-\hat{y}_i)^2 \quad \mbox{and} \quad AME=\max_{_i=1,\ldots, m} |y_i-\hat{y}_i|,$$
where $y_i$ is the true value of computer code at input $\boldsymbol{x}_i$ from the test data and $\hat{y}_i$ is the prediction of the GP at the same input.  Both of these measures are used to assess the overall model performance. 
%%where the sample size $n=10d$.  

\begin{figure}[!h]
\begin{knitrout}
\definecolor{shadecolor}{rgb}{0.969, 0.969, 0.969}\color{fgcolor}
\includegraphics[width=\linewidth]{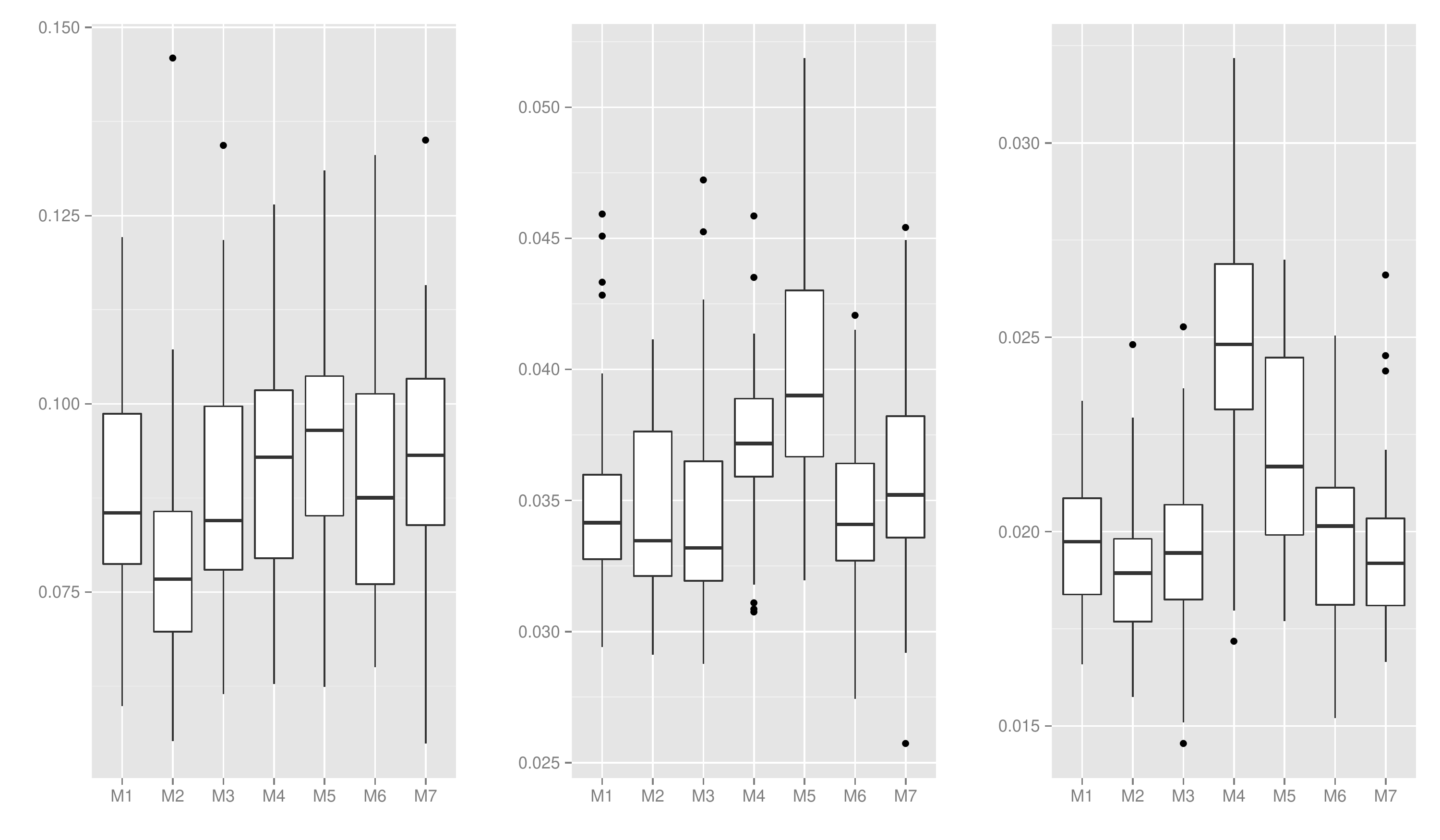} 

\end{knitrout}
\caption{Box plots of RMSE values for one function with sample sizes 50 (left), 100 (middle), and 150 (right) fit by each of the seven methods.}
\label{fig:boxplots}
\end{figure}

\cite{CheLoeSac2015} discuss the importance of having extensive simulations and multiple data sets to validate a new statistical procedure.  This working example and an incomplete list of other papers \citep{LoeSacWel2009,LoeMooWil2010,WilLoeMoo2011, GraLee2012,GraApl2015, ButHayHum2014,LamNot2008} all have a similar structure of a reasonably comprehensive set of simulations and number of methods that need to be assessed.  The papers above typically show the results as a series of side-by-side boxplots  or 
side-by-side dotcharts for each method, with one plot for each test function and sample size.  
Conclusions are then drawn from looking at a handful of boxplots which often look very cluttered and usually do not provide clear evidence as to the best method(s).  Alternatively, the results will be summarized in a table of average performance for each method separated by function and sample size.  These tables are usually overwhelming to look at and interpretations are incredibly inefficient.  Using our working example, the plot in Figure \ref{fig:boxplots}  illustrates this currently widely used approach.

Figure \ref{fig:boxplots} shows the boxplots for RMSE values corresponding to run sizes of 50, 100, and 150 respectively for one of the 24 test functions under consideration.  Within each plot, the seven boxes correspond to the seven different methods used to fit the data. Note that the scales of the plots do not correspond. The RMSE values in the first plot are much larger than that of the second and the same can be seen between the second and third. If we were to plot them all on the same $y$-axis, the details of the second and third plots would not be clearly seen. Although these plots certainly provide information on the fit of the seven models, it is not possible to determine which method(s) are performing best over all. As an example, for the middle plot, cases could be made to support methods 2, 3 and 6.  Furthermore, we would typically be looking at similar plots for multiple test functions and trying to draw conclusions from say ten or twelve plots similar to just the one shown. In our opinion, these plots, just like many of the plots in the literature, do not provide a clear picture of the evidence for or against a method.  

In the papers above and other work, we find the results of various methods to appear extremely similar in box or dot plots and tables and it is nearly impossible to choose a clear winner for a single function, let alone trying to choose a clear winner overall.  We find the lack of an easy way to visualize the results from these simulations to be a large problem. We find the goal for using these boxplots (or dotcharts and tables), which is to provide overall evidence as to the usefulness of a chosen procedure, is not being met.    

The need for better summaries and the need for better designed simulation studies is of course not new.  \cite{Chi2015} discussed the need for statistically analyzing simulation results and even using principals of experimental design to reduce the number of simulations being preformed.  We agree with the approach, however, we focus attention on the need for simple graphical methods that allow for quick assessments of what is practically significant.  In many simulation studies, the number of replicates can be very high which will often lead to very small residual errors which might unduly inflate the significance of the results.  In what follows we discuss a very simple graphical method to assess which methods are expected to do well in practice. 

\section{A Simple Minded Approach  \label{sect:sm}} 

Motivated by benchmarking in the optimization literature, we seek a simple tool that allows one to quickly assess the overall  performance of a set of methods.  In many simulation studies, there is a primary factor of interest and multiple other factors are introduced to simply create more experimental cases to be tested. Additionally, these are often replicated a large number of times.  The other factors and the replicates in this case are nuisance variables that are not of primary interest. Many of the common plots of course display results for a fixed level of one of these factors and only show replicates and methods in a single plot such as those seen in Figure \ref{fig:10dplot}.  This leaves the reader, and the authors in many cases, completely baffled as to what method is a clear winner.  In our opinion the reasons for this are often two fold. First, the graphical (or tabular) displays used are simply inadequate and secondly, we often have inadequate methods for judging performance.  Although optimization benchmarking deals with very different problems, many of the arguments made in the benchmarking literature can be directly applied to assessing the quality of a statistical method. 

Benchmarking in optimization attempts to ask the question of which optimization routine (solver) is the best overall at optimizing a large class of test problems \citep{MorWil2009}.  This involves having a very large test set of problems to compare and then running each numerical solver on each of these test problems.  For a set of $p=1, \ldots, P$ problems and $s=1, \ldots, S$ solvers the goal is to find solver $s^*$ which is the ``best'' at solving all problems.  The hope of course is that the best solver on this test set will also be the best solver on a set of untested problems.   Given an appropriate test set of functions we do not view this as being contradictory to the goals of assessing a statistical method.    

Two common plots in Benchmarking that warrant some discussion are performance profiles and data profiles \citep{MorWil2009}.  Performance profiles define the measure of interest for each solver to be the time taken to solve a given problem.  Each solver is assessed by comparing the ECDFs of the time taken to solve all problems.  On the other hand, data profiles display the proportion of problems solved by a given method for a fixed time budget.  Several modifications of these plots also exist that display the proportion of problems solved within a given accuracy.  In order for either of these plots to be effective it is important to have a standardized measure of performance. For example, in performance profiles,  CPU time is typically used as opposed to the number of function evaluations since some functions might take longer to evaluate then others and some algorithms run longer then others.  When displaying accuracy it is typical to compute a relative error for a problem which is given by $F_p=(f_p-f^*)/f^*$ where $f^*$ is the known solution and $f_p$ is the solution found by the solver.  When the solution $f^*$ is unknown it is common to replace $f^*$ with the best solution found by all solvers.  Standardizing the results allows for easy use of the ECDF to quickly assess which solver is best for a particular problem.

Drawing the parallels between statistical problems and benchmarking, most statistical simulation studies consist of $m=1, \ldots, M$ methods that need to be compared. The comparisons are done over a wide array of factors that represent possible scenarios of interest and typically many replicates, all of these multiple combinations could be viewed as the set of $p=1,\ldots, P$ problems where the comparisons are performed.  Many papers tend to do a separate analysis of each of the $P$ problems and then try to draw general conclusions on the appropriateness of the given method. This of course is often difficult and results in an overwhelming number of plots or tables to be compared.  In order to avoid these difficulties and following the procedures from benchmarking, the simple minded approach to choose the best method would be to simply plot the ECDF of a suitable measure of accuracy for each of the $P$ problems that were analyzed.  There are two perceived issues with this approach.  Firstly, the suggestion is to collapse over all of the different problems $P$ which results in a perceived loss of any functional information.  In what follows below, we show that one does not need to collapse over all problems.  Additionally, we show that significant benefit is gained by collapsing.  This benefit is amplified when one takes the view point of choosing a method that is expected to perform well on a variety of problems.  The second concern is defining a suitable measure of accuracy.  One obvious choice is the standardized measured $F_p$ shown above.  In the next section, we illustrate the use of this tool on the data from the motivating example and discuss an alternative measure of standardizing the results.   

\section{Assessing Statistical Methods \label{sect:asm}} 

In order to define a suitable measure of accuracy that is comparable across all of the different problems we consider two approaches.  For now, we will focus on using either the RMSE or AME obtained by fitting the GP for each of $7$ different methods and divide that value by the RMSE or AME from the model using the the trivial predictor of $\bar{y}$.  This allows us to easily plot very different RMSE or AME values on the same plot and also allows for a quick assessment of quality since values of this ratio larger than say 0.5 would be classified as an extremely poor fitting GP.  Since these measures of accuracy are all relatively close to one another, we typically take a log base 10 transformation to induce some spread in the results. Other transformations could be used including a natural log, but we use log base 10 since this provides the negative value of the decimals of accuracy achieved by the method. 

\begin{figure}[!h]
\begin{knitrout}
\definecolor{shadecolor}{rgb}{0.969, 0.969, 0.969}\color{fgcolor}
\includegraphics[width=\linewidth]{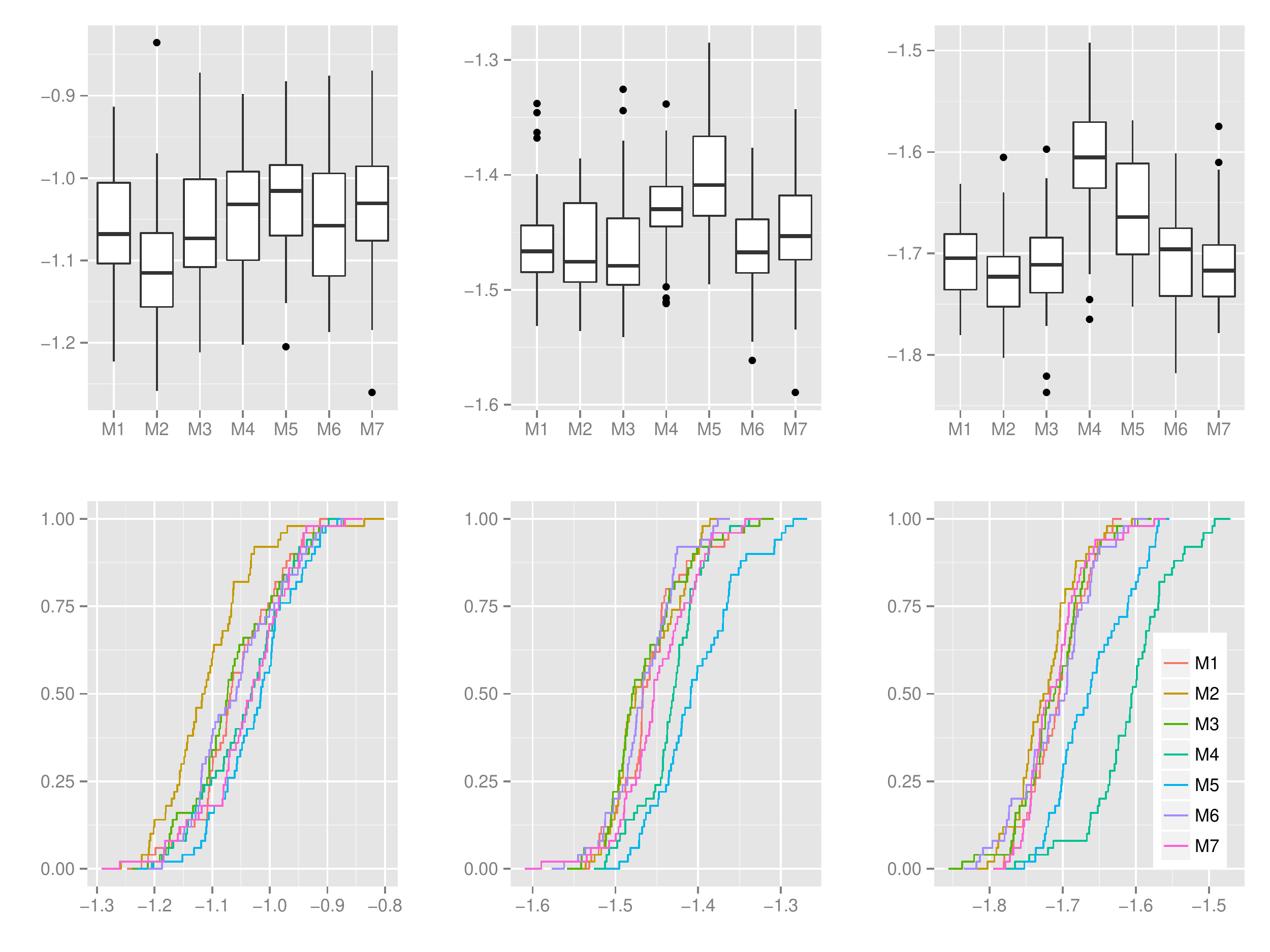} 

\end{knitrout}
\caption{Box plots and ECDFs for seven methods on one test function for sample sizes 50, 100, and 150 from left to right.}
\label{fig:box_ecdf}
\end{figure}

The plot in Figure \ref{fig:box_ecdf} shows the log 10 transformed RMSE values displayed as both boxplots in the top row and as the ECDFs in the bottom row for the same data displayed in Figure \ref{fig:boxplots}.  Note that using logged RMSE values in the boxplot still results in multiple confusing plots that do not allow a clear winner to be identified.  For the ECDFs since we are taking the log of values less than one, we end up with negative values. The more negative the value, the smaller the RMSE. Therefore, the line that sits farthest left represents the method that performs the best on the problems. We can clearly see that for sample sizes 50 and 150 (the left and right plots) method 2 performs best. For sample size 100, it is not as clear. Method 3 is the best for mid range RMSE values, but as RMSE values increase, methods 2 and 6 both outperform 3. We argue that the tail end of the ECDF is most important. Differences in more negative values correspond to small differences in RMSE, while the larger values correspond to more noticeable differences in performance at a point where performance is hardly better, if any better, than the trivial predictor. For this reason, methods 2 and 6 are more attractive than 3. Clearly, we can see a lot more about the performance of the methods in the ECDFs as compared to the minimal information provided by boxplots. We can very quickly see that methods 4 and 5 are not performing well at all compared to the other methods and that method 2 is consistently performing well out of the seven.

In the case above, we considered one problem and separate plots for three different sample sizes. We can also collapse the three sample sizes into one plot as in Figure \ref{collapse}. It is still clear to see which method is doing best overall. Additionally, the approximate locations in the plot where RMSE values shift from one run size to the next should be noticeable upon visual inspection. If it is not easy to identify these locations, vertical lines can be added in the plot to show roughly where these boundaries are. Although not needed in this case, we have included such vertical lines. The advantage of this plot over the set of ECDF plots above is that all of the information is clearly condensed into one display. At once, we can see if there is one clear winner or if for some run sizes the winner is not as obvious.  From the plots it is easy to see that methods 4 and 5, the cosine transformed  designs \citep{DetPep2010}, get progressively worse as the sample size increases.  Additionally we see that method 2, the maximin LHS, does much better at smaller sample sizes, which is perhaps not that surprising since it is focused on space filling which will become less important as the density of points increases.

\begin{figure}[!h]
\begin{knitrout}
\definecolor{shadecolor}{rgb}{0.969, 0.969, 0.969}\color{fgcolor}
\includegraphics[width=\linewidth]{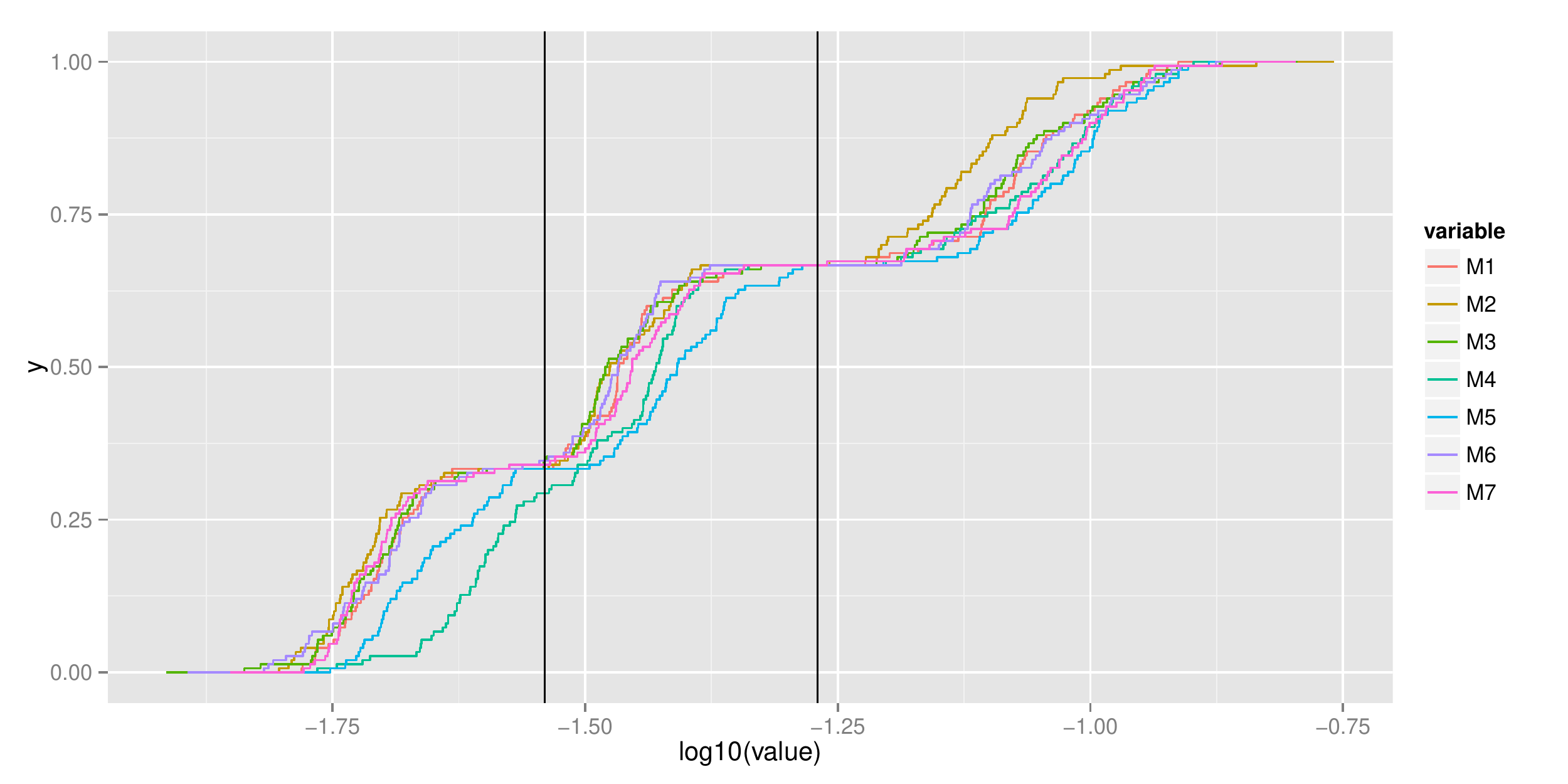} 

\end{knitrout}
\caption{RMSE values for one function collapsed over sample sizes 50, 100, and 150 with approximate boundaries indicated by vertical lines.}
\label{fig:collapse}
\end{figure}

\begin{figure}[!h]
\begin{knitrout}
\definecolor{shadecolor}{rgb}{0.969, 0.969, 0.969}\color{fgcolor}
\includegraphics[width=\linewidth]{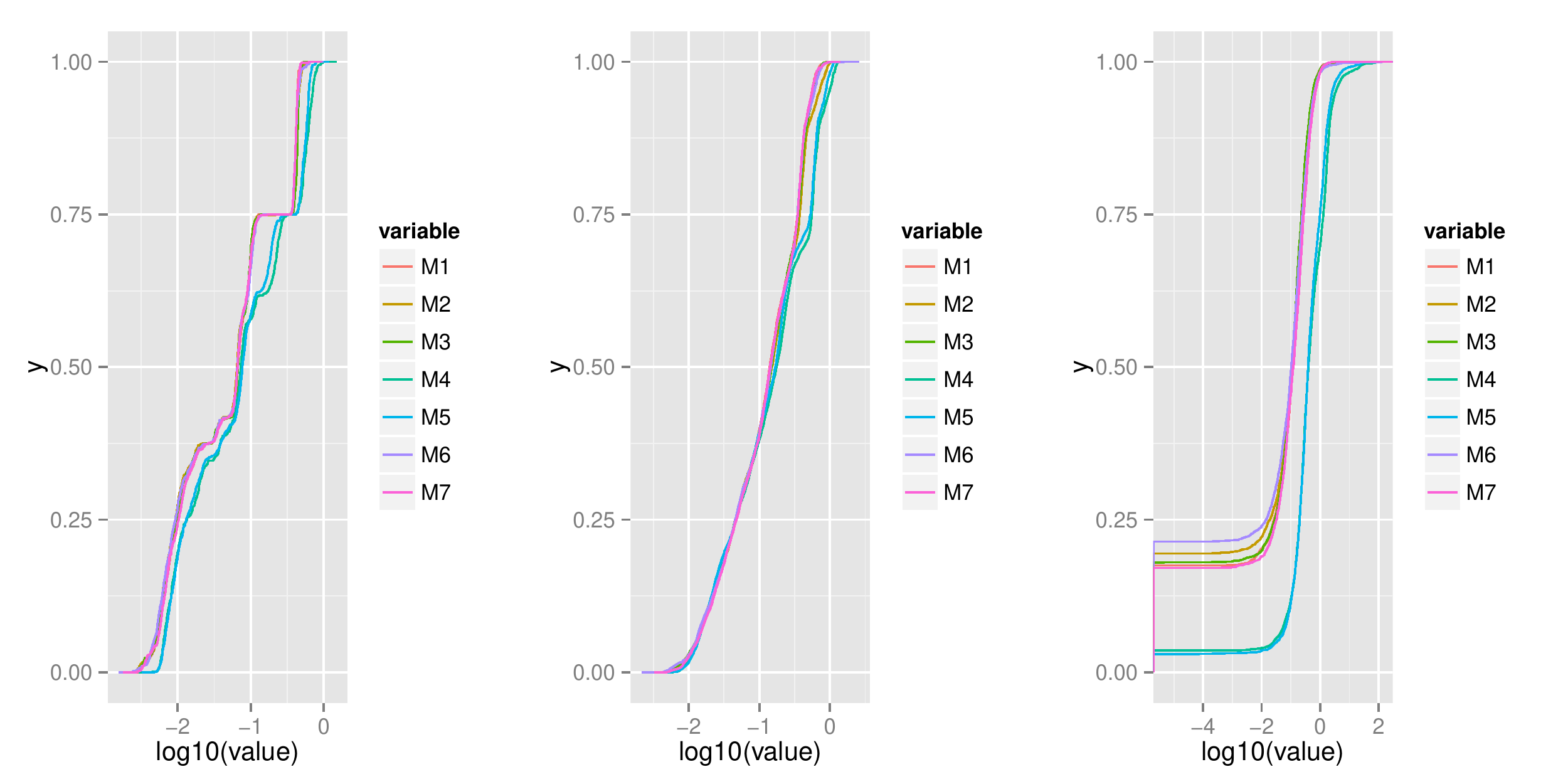} 

\end{knitrout}
\caption{ECDFs for RMSE values collapsed over all 10d simulations (left plot), AME values collapsed over 10d simulations (middle), and alternate standardized RMSE values for 10d simulations (right).}
\label{fig:10dplot}
\end{figure}

Finally, we can also collapse over many functions at once. The left plot of Figure \ref{fig:10dplot} includes RMSE values for all 24 test functions evaluated for sample size 10d. Therefore, we have 24 problems each with 50 replicates evaluated by each of the seven methods. We notice that overall Methods 1, 2, 3, 6, and 7 are performing about equally well. The middle plot of Figure \ref{fig:10dplot} again uses all 24 problems but shows standardized AME values for each of the 50 replicates rather than RMSE data. We draw similar conclusions from this plot supporting those from the RMSE data.

Until now, we have been standardizing RMSE and AME values by the corresponding RMSE and AME values obtained by fitting the trivial predictor. It is important to note that depending on the problem, the standardization used may differ. To produce useful ECDFs, care must be taken in this step. We present here an alternate form of standardizing for our problem set to show another possibility and also to see that we obtain consistent results. The right plot in Figure \ref{fig:10dplot} uses the standardization $F_p=(f_p-f^*)/f^*$, where $f^*$ is the lowest RMSE value from each of the seven methods for each replicate. Again, we see similar results between this and the other two plots, although this form of standardization shows a larger discrepancy in performance of Methods 4 and 5 compared to the rest. The advantage of this plot may be that we can see for what percentage of the problems each method has the best solution. For example, we see that Method 6 has the best solution nearly 25\% of the time.

\section{Concluding Remarks \label{sect:cr}} 

In this paper we have introduced a very simple graphical method to assess the performance of a given statistical procedure.  We have additionally argued that in many cases significant benefits are to be gained by plotting multiple problems on the same graph so that one can quickly and reliably decide which methods are being outperformed on a large class of problems. Compared to plotting boxplots or dotplots for all of the results of one method, the ECDF has the advantage of displaying every single data point in two-dimensions and allows for an easy interpretation. One can read off what percentage of problems have been solved within a given accuracy.  When this accuracy is converted to standardized performance measure, it also allows one to see which method results in the best fit over all of the different problems.  Although the ECDF is a very simple tool, it is far more sophisticated than a laundry list of tables or pages of boxplots or dotplots displaying results for one test at a time. In our opinion, displays of this variety do a significantly better job at obfuscating the relevant information and leaving the authors and researchers confused rather than indicating which method preforms best.  

The ECDF allows for quick assessments of methods over a large array of problems to get an overall view while of course not precluding comparisons on individual functions.  The major advantage of the ECDF is its ability to quickly show similarities and differences that are not as easy to see with dotplots or boxplots which present themselves as being cluttered.   We hope that readers of this paper agree with our opinions and strongly encourage everyone to rely on the ECDF, at least as a starting point, to display relevant statistical information from simulations.

%
%
%The method allows one to show all of the data in very simple format, allows one to plot multiple methods on one plot to get a general overall picture. It definitely provides a much better tools than a laundry list of tables or a never ending display of graphs that do a better job of obfuscating the information than allowing the user to make a quick assessment, while also allowing for more subtle comparisons of results.  
%
%

\centerline{\bf ACKNOWLEDGEMENTS}

\renewcommand{\baselinestretch}{1.0} The research of Loeppky was supported by  Natural Sciences and Engineering Research Council of Canada Discovery Grant  ({RGPIN-2015-03895 }). The research of Arnold was conducted at the University of British Columbia with the support of a Natural Sciences and Engineering Research Council of Canada undergraduate research award. 

%\medskip

\baselineskip 0.7\normalbaselineskip
\bibliographystyle{asa}
\bibliography{psn}

\end{document}